\documentclass[russian]{jetpl}

\twocolumn

\usepackage{amsmath,amssymb,color,pstricks,bm,alltt}

\usepackage[T1]{fontenc}
\usepackage[cp1251]{inputenc}  
\usepackage{babel}
\usepackage{graphicx}

\begin{document}

\lat

\title{Temperature of superconducting transition for very strong coupling
in antiadiabatic limit of Eliashberg equations}

\rtitle{Temperature of superconducting transition
in antiadiabatic limit of Eliashberg equations}

\sodtitle{Temperature of superconducting transition for very strong coupling
in antiadiabatic limit of Eliashberg equations}

\author{M.\ V.\ Sadovskii\thanks{E-mail: sadovski@iep.uran.ru},
}

\rauthor{M.\ V.\ Sadovskii}

\sodauthor{M.\ V.\ Sadovskii}

\address{Institute for Electrophysics, Russian Academy of Sciences,
Ural Branch, Amundsen str. 106, Ekaterinburg 620016, 
Russia
}


\abstract{
It is shown that the famous Allen -- Dynes asymtotic limit for superconducting
transition temperature in very strong coupling region
$T_{c}>\frac{1}{2\pi}\sqrt{\lambda}\Omega_0$ (where $\lambda\gg 1$
-- is Eliashberg -- McMillan electron -- phonon coupling constant
and $\Omega_0$ -- the characteristic frequency of phonons) in antiadiabatic limit 
of Eliashberg equations $\Omega_0/D\gg 1$ ($D\sim E_F$ is conduction band
half-width and $E_F$ is Fermi energy) is replaced by 
$T_c>(2\pi^4)^{-1/3}(\lambda D\Omega_0^2)^{1/3}$, with the upper limit
for $T_c$ given by $T_c<\frac{2}{\pi^2}\lambda D$.
}


\maketitle


\section{Introduction}

The discovery of superconductivity \cite{H3S} with critical temperature up to
$T_c=$ 203 in pressure interval of 100-250 GPa (in diamond anvils) in H$_3$S 
system initiated the flow of articles with experimental studies of
high -- temperature superconductivity of hydrides in megabar region
(cf. reviews \cite{Er,ErD}). Theoretical analysis immediately confirmed that
these record -- breaking values of $T_c$ are ensured by traditional electron --
phonon interaction in the limit of strong -- enough electron -- phonon
coupling \cite{Ash,Grk-Krs}. More so, the detailed calculations performed for
quite a number of hydrides of transition metals under pressure \cite{Ash} lead
to prediction of pretty large number of such systems with record values of $T_c$.
In some cases these predictions were almost immediately confirmed by experiment, 
in particular the record values of $T_c=$ 160--260 K were achieved in
in LaH$_{10}$ \cite{DrEr,Som}, ThH$_{10}$
\cite{P1}, YH$_{6}$ \cite{P2}, (La,Y)H$_{6,10}$ \cite{P3}.
At last, some time ago the psychological
barrier was overpassed, when in Ref. \cite{RT} superconductivity was obtained
with $T_c=$ 287.7$\pm$1.2 K (i.e. near +15 degrees of Celsius) in C-H-S system
at pressure of 267$\pm$10 GPa.

The principal achievement of these works was, before everything else, the
demonstration of absence of any significant limitations for $T_c$, within the
traditional picture of electron -- phonon mechanism of Cooper pairing,
contrary to a common opinion that $T_c$ due to it can not exceed 30-40 K. 
Correspondingly, even more demanding now is the problem of the upper limit
of $T_c$ values, which can be achieved with this mechanism of pairing.

Since BCS theory appeared it became obvious the the increase of $T_c$
can be achieved either by the increase of the frequency of phonons,
responsible for Cooper pairing, or by the increase of the effective
interaction of these phonons with electrons. 
These problems were thoroughly studied by different authors. The most
developed approach to description of superconductivity in electron --
phonon system is Eliashberg -- McMillan theory  \cite{Grk-Krs,Scal,All}.
It is well known that this theory is entirely based on the applicability
of adiabatic approximation and Migdal theorem \cite{Mig}, which allows to
neglect vertex corrections while calculating the effects of electron --
phonon interactions in typical metals. The actual small parameter of
perturbation theory in these calculations is
$\lambda\frac{\Omega_0}{E_F}\ll 1$, where $\lambda$ is the dimensionless
coupling constant of electron -- phonon interactions, $\Omega_0$ is 
characteristic frequency of phonons and $E_F$ is Fermi energy of electrons. 
In particular, this means that vertex corrections in this theory can be
neglected even in case of $\lambda > 1$, as we always have an inequality
$\frac{\Omega_0}{E_F}\ll 1$ valid for typical metals.

In recent papers \cite{Sad_18,Sad_19,Sad_20} we have shown that in case of
strong nonadiabaticity, when $\Omega_0\gg E_F$, a new small parameter appears 
in the theory
$\lambda_D\sim \lambda\frac{E_F}{\Omega_0}\sim\lambda\frac{D}{\Omega_0}
\ll 1$ ($D$ is electronic band half-width), so that corrections to
electronic spectrum become irrelevant. Vertex corrections can also be 
neglected, as it was shown in an earlier Ref. \cite{Ikeda}.
In general case the renormalization of electronic spectrum (effective mass
of an electron) is determined by a new dimensionless constant
$\tilde\lambda$, which reduces to the usual $\lambda$ in the adiabatic limit,
while in strong antiadiabatic limit it tends to $\lambda_D$. 
At the same time, the temperature of superconducting transition $T_c$ in 
antiadiabatic limit is determined by Eliasnberg -- McMillan pairing constant
$\lambda$, generalized by the account of finite phonon frequencies.

For the case of interaction with a single optical (Einstein) phonon in
Ref. \cite{Sad_18} we have obtained the single expression for
$T_c$, which is valid both in adiabatic and antiadiabatic regimes and smoothly
interpolating in between:
\begin{equation}
T_c\sim 
\frac{D}{1+\frac{D}{\Omega_0}}\exp\left(-\frac{1+\tilde\lambda}
{\lambda}\right)
\label{Tc_opt_single}
\end{equation}
where $\tilde\lambda=\lambda\frac{D}{\Omega_0+D}$ is smoothly changing from
$\lambda$ for $\Omega_0\ll D\sim E_F$ to $\lambda_D$ in the limit of 
$\Omega_0\gg D\sim E_F$.

Besides the questions related to possible limits of $T_c$ in hydrides, 
where possibly some small pockets of the Fermi surface with small Fermi
energies exist \cite{Grk-Krs}, the interest to the problem of superconductivity
in strongly antiadiabatic limit is stimulated by the discovery of a number of
other superconductors, where adiabatic approximation can not be considered
valid and characteristic phonon frequencies is of the order or even exceed
the Fermi energy of electrons. Typical in this respect are intercalated
systems with monolayers of FeSe, and monolayers of FeSe on substrates like
Sr(Ba)TiO$_3$ (FeSe/STO) \cite{UFN}. With respect to FeSe/STO this was first
noted by Gor'kov \cite{Gork_1,Gork_2}, while discussing the idea of the
possible mechanism of increasing superconducting transition temperature $T_c$ 
in FeSe/STO due to interactions with high- energy optical phonons of
SrTiO$_3$ \cite{UFN}. Similar situation appears also in an old problem of
superconductivity in doped SrTiO$_3$ \cite{Gork_3}.

\section{Limits for superconducting transition temperature in case of
very strong electron -- phonon coupling}

The general equations of Eliashberg -- McMillan theory determining superconducting gap 
$\Delta(\omega_n)$ in Matsubara representation $(\omega_n=(2n+1)\pi T)$ can be written as 
\cite{Grk-Krs,Scal,All}:
\begin{eqnarray}
\Delta(\omega_n)Z(\omega_n)=T\sum_{n'}\int_{-D}^{D}d\xi\int_{0}^{\infty}
d\omega\alpha^2(\omega)F(\omega)\times\nonumber\\
\times D(\omega_n-\omega_{n'};\omega)\frac{\Delta(\omega_n')}
{\omega^2{_n'}+\xi^2+\Delta^2(\omega_{n'})}
\label{El_Mats}
\end{eqnarray}
\begin{eqnarray}
Z(\omega_n)=1+\frac{\pi T}{\omega_n}\sum_{n'}
\int_{-D}^{D}d\xi\int_{0}^{\infty}d\omega
\alpha^2(\omega)F(\omega)\times\nonumber\\
\times D(\omega_n-\omega_{n'};\omega)
\frac{\omega_n'}
{\omega^2{_n'}+\xi^2+\Delta^2(\omega_{n'})}
\label{Zgen}
\end{eqnarray}
where we have introduced
\begin{equation}
D(\omega_n-\omega_{n'};\omega)=\frac{2\omega}{(\omega_n-\omega_{n'})^2+\omega^2}
\label{Dw}
\end{equation}
Here $\alpha^2(\omega)F(\omega)$ is McMillan's function, $F(\omega)$ is the
phonon density of states, and for simplicity we assume here the model of
half-filled band of electrons with finite width $2D$ ($D\sim E_F$) with constant
density of states (two -- dimensional case). 

We also neglect here the effects of Coulomb repulsion leading to the
appearance of Coulomb pseudopotential $\mu^{\star}$, which is usually small and
more or less irrelevant in the region of very strong electron -- phonon
attraction \cite{Grk-Krs,Scal,All}.

Then, taking into account:
\begin{eqnarray}
\int_{-D}^{D}d\xi\frac{1}{\omega_{n'}^2+\xi^2+\Delta^2(\omega_{n'})}=\nonumber\\
=\frac{2}{\sqrt{\omega_{n'}^2+\Delta^2(\omega_{n'})}}
arctg\frac{D}{\sqrt{\omega_{n'}^2+\Delta^2(\omega_{n'})}}\to\nonumber\\
\to\frac{2}{|\omega_{n'}|}
arctg\frac{D}{|\omega_{n'}|} \ \mbox{при}\ \Delta(\omega_{n'})\to 0
\label{int_xii}
\end{eqnarray}
the linearized Eliashberg equations take the following general form:
\begin{eqnarray}
\Delta(\omega_n)Z(\omega_n)=T\sum_{n'}\int_{0}^{\infty}d\omega\alpha^2(\omega)
F(\omega)\times\nonumber\\
\times D(\omega_n-\omega_{n'};\omega)
\frac{2\Delta(\omega_{n'})}{|\omega_{n'}|}
arctg\frac{D}{|\omega_{n'}|}
\label{lin_Delta_gen}
\end{eqnarray}
\begin{eqnarray}
Z(\omega_n)=1+\frac{T}{\omega_n}\sum_{n'}\int_{0}^{\infty}d\omega
\alpha^2(\omega)F(\omega)\times\nonumber\\
\times D(\omega_n-\omega_{n'};\omega)
\frac{\omega_{n'}}{|\omega_{n'}|}
2 arctg\frac{D}{|\omega_{n'}|}
\label{lin_Z_gen}
\end{eqnarray}
Consider the equation for $n=0$ determining   
$\Delta(0)\equiv\Delta(\pi T)=\Delta(-\pi T)$, which follows directly from
Eqs. (\ref{lin_Delta_gen}), (\ref{lin_Z_gen}):
\begin{eqnarray}
\Delta(0)=T\sum_{n'\neq 0}\int_{0}^{\infty}d\omega\alpha^2(\omega)F(\omega)
\frac{2\omega}{(\pi T-\omega_{n'})^2+\omega^2}\times\nonumber\\
\times\frac{2\Delta(\omega_{n'})}{|\omega_{n'}|}
arctg\frac{D}{|\omega_{n'}|}
\label{D0_eq}
\end{eqnarray}
Leaving in the r.h.s. only the contribution from $n'=-1$, we immediately
obtain the {\em inequality}:
\begin{equation}
1>\frac{2}{\pi}\int_{0}^{\infty}d\omega\alpha^2(\omega)F(\omega)
\frac{2\omega}{(2\pi T)^2+\omega^2}arctg\frac{D}{\pi T}
\label{AD_inq_gen}
\end{equation}
which generalizes the similar inequality first obtained in Allen -- Dynes
paper \cite{AD} and determining the {\em lower} boundary for $T_c$.
For Einstein model of phonon spectrum we have $F(\omega)=\delta(\omega-\Omega_0)$,
so that Eq. (\ref{AD_inq_gen}) is reduced to:
\begin{equation}
1>\frac{2}{\pi}\lambda arctg\frac{D}{\pi T}\frac{\Omega^2_0}{(2\pi T)^2+\Omega^2_0}
\label{AD_gen_ineq}
\end{equation}
where $\lambda=2\alpha^2(\Omega_0)/\Omega_0$ is dimensionless pairing coupling
constant. 
For $D\gg\pi T$ we immediately obtain the Allen -- Dynes result \cite{AD}:
\begin{equation}
T_c>\frac{1}{2\pi}\sqrt{\lambda-1}\Omega_0\to 0.16\sqrt{\lambda}\Omega_0\ \mbox{при}\ \lambda\gg 1\
\label{T_c_AD}
\end{equation}
which in fact determines the asymptotic behavior of $T_c$ in the region of very
strong coupling $\lambda\gg 1$. The exact numerical solution of Eliashberg equation
\cite{AD} produces for $T_c$ the result like (\ref{T_c_AD}) with replacement of
numerical coefficient 0.16 by 0.18. This asymptotic behavior rather satisfactory
describes the values of $T_c$ already for $\lambda>2$.

In the case of general phonon spectrum it is sufficient to replace here
$\Omega_0\to\langle\Omega^2\rangle^{1/2}$, where
\begin{equation}
\langle\Omega^2\rangle=\frac{2}{\lambda}\int_{0}^{\infty}d\omega
\alpha^2(\omega)F(\omega)\omega
\label{sq_freq}
\end{equation}
is the average (over the spectrum) square frequency of phonons, and the
general expression for the coupling constant is \cite{Grk-Krs,Scal,All}:
\begin{equation}
\lambda=2\int_{0}^{\infty}\frac{d\omega}{\omega}\alpha^2(\omega)F(\omega)
\label{lamb_Eli}
\end{equation}
For $D\ll\pi T$ from Eq. (\ref{AD_gen_ineq}) we obtain
\begin{equation}
T>\frac{1}{2\pi}\sqrt{\lambda^*(T)-1}\Omega_0
\label{TcAD-star}
\end{equation}
where
\begin{equation}
\lambda^*(T)=\frac{2D}{\pi^2 T}\lambda
\label{lamb-star}
\end{equation}
so that in strongly antiadiabatic limit we get:
\begin{equation}
T_c>(2\pi^4)^{-1/3}(\lambda D\Omega_0^2)^{1/3}\approx 0.17(\lambda D\Omega_0^2)^{1/3}
\label{Tc-AD-anti}
\end{equation}
From the obvious requirement of $\lambda^*(T)>0$ we obtain the condition:
\begin{equation}
T_c<\frac{2}{\pi^2}\lambda D
\label{ineq_T}
\end{equation}
which limits $T_c$ from the above.
\begin{figure}
\includegraphics[clip=true,width=0.5\textwidth]{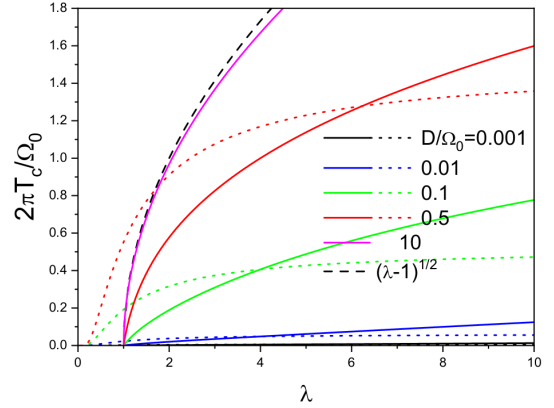}
\caption{Fig. 1. Temperature of superconducting transition in Einstein model of
phonon spectrum in units of $2\pi T_c/\Omega_0$, as function of pairing constant
$\lambda$ for different values of the inverse adiabaticity parameter
$\frac{D}{\Omega_0}$.
Dotted lines show the dependencies for $2\pi T_c/\Omega_0$ in the region of weak
and intermediate couplings (\ref{Tc_opt_single})
\cite{Sad_18}. Black dashed line --- Allen -- Dynes estimate valid in adiabatic
limit \cite{AD}}
\label{TcAD_anti_gen}
\end{figure}
\begin{figure}
\includegraphics[clip=true,width=0.5\textwidth]{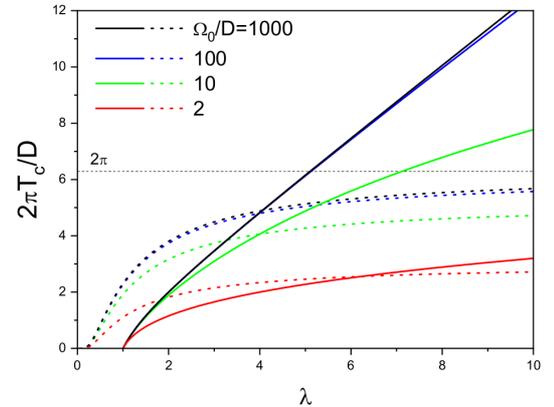}
\caption{Fig. 2. Temperature of superconducting transition in Einstein model of
phonon spectrum in units of $2\pi T_c/D$, as function of pairing constant
$\lambda$ fro different values of adiabaticity parameter $\frac{\Omega_0}{D}$.
Dotted lines show the dependencies for $2\pi T_c/D$ in the region of weak and
intermediate couplings (\ref{Tc_opt_single}) \cite{Sad_18}. Black line --- the
limiting behavior given by $\frac{2}{\pi^2}\lambda D$}
\label{TcAD_anti_gen_D}
\end{figure}

Thus we require the validity of the following inequality:
\begin{equation}
(2\pi^4)^{-1/3}(\lambda D\Omega_0^2)^{1/3}<T_c<\frac{2}{\pi^2}\lambda D
\label{doubl_ineq}
\end{equation}
which reduces to:
\begin{equation}
\Omega_0<\frac{4}{\pi}\lambda D\approx 1.27\lambda D\ \mbox{или}\
\frac{D}{\Omega_0}>\frac{0.78}{\lambda}
\label{ineq_D_lam}
\end{equation}
so that for our analysis to be self -- consistent it is required to have: 
\begin{equation}
\lambda\gg\frac{\Omega_0}{D}\gg 1
\label{lamb_big}
\end{equation}
where the last inequality corresponds to strong antiadiabatic limit. 
Correspondingly, all the previous estimates are not valid for
$\lambda\sim 1$ and can only describe the limit of very strong coupling.

In Fig.1 and Fig. 2 we show the results of numerical comparison of the 
boundaries for $T_c$, following from Eq. (\ref{AD_gen_ineq}) with the values of
transition temperature in the region of weak and intermediate coupling
following from Eq. (\ref{Tc_opt_single}), for different values of adiabaticity
parameter $\Omega_0/D$. It is clear that in the vicinity of intersections of
dotted and continuous lines on these graphs we actually have the smooth
crossover from $T_c$ behavior in the region of weak and intermediate
coupling to its asymptotic behavior in the region of very strong coupling
$\lambda\gg 1$. It is also seen that the increase of phonon frequencies and
crossover to antiadiabatic limit does not lead, in general, to the increase
of $T_c$ as compared with adiabatic case.

\section{Conclusions}

In this paper we have considered the case of very strong electron -- phonon
coupling in Eliashberg -- McMillan theory, including the antiadiabatic
situation with phonons of very high frequency (exceeding the Fermi energy $E_F$). 
The value of mass renormalization is in general determined by the coupling
constant $\tilde\lambda$ \cite{Sad_18}, which is small in antiadiabatic limit.
At the same time, the pairing interaction is always determined by the
standard coupling constant $\lambda$ of Eliashberg -- McMillan theory, 
appropriately generalized by taking into account the finite values of
phonon frequencies \cite{Sad_18}. However, the simplest estimates
\cite{Sad_18,Sad_20} show, that in antiadiabatic situation this constant in general
rather rapidly drops with the growth of phonon frequency 
$\Omega_0$ for $\Omega_0\gg E_F$. In this sense, the asymtotics of $T_c$ for
very strong coupling, discussed above, can be possibly achieved only in some 
exceptional cases. Even in this case, as it is clear from our results, the
transition into antiadiabatic region can not  increase $T_c$ as compared with
the standard adiabatic situation.

While the usual expression for $T_c$ in terms of pairing constant  $\lambda$ 
and characteristic phonon frequency $\Omega_0\sim\langle\Omega^2\rangle^{1/2}$ 
are quite convenient and clear, it is to be taken into account that these
parameters are in fact not independent. As it is seen from expressions like
(\ref{sq_freq}) and (\ref{lamb_Eli}), these parameters are determined by the
same Eliashberg -- McMillan function $\alpha^2(\omega)F(\omega)$. 
Correspondingly, there are limitations for free changes of these parameters
in estimates of optimal (maximal) values of $T_c$.

The author is grateful to E.Z. Kuchinskii for discussions and help with
numerical computations. This work is partially supported by RFBR grant
No. 20-02-00011.


\begin{thebibliography}{99}

\bibitem{H3S}A.P. Drozdov, M.I. Eremets, I.A. Troyan, V. Ksenofontov,
S.I. Shylin. Nature {\bf 525}, 73 (2015)
\bibitem{Er}M.I. Eremets, A.P. Drozdov. Usp. Fiz. Nauk {\bf 186}, 1257 (2016)
\bibitem{ErD}C.J. Pickard, I. Errea, M.I. Eremets. Annual Reviews of Condensed Matter Physics
{\bf 11}, 57 (2020)
\bibitem{Ash}H. Liu, I.I. Naumov, R. Hoffman, N.W. Ashcroft, R.J. Hemley. PNAS {\bf 114},
6990 (2018)
\bibitem{Grk-Krs}L.P. Gor'kov, V.Z. Kresin. Rev. Mod. Phys. {\bf 90}, 01001 (2018)
\bibitem{DrEr}A.P. Drozdov, et al.  Nature {\bf 569}, 528 (2019)
\bibitem{Som}M. Somayazulu, et al. Phys. Rev. Lett. {\bf 122}, 027001 (2019)
\bibitem{P1}D.V. Semenok et al. Materials Today {\bf 33}, 36 (2020)
\bibitem{P2}I.A. Troyan et al. ArXiv:1908.01534; Advanced Materials (2021)
\bibitem{P3}D.V. Semenok et al. ArXiv:2012.04787; Nature Materials (2021)
\bibitem{RT}E. Snider, N. Dasenbrock-Gammon, R. McBride, M. Debessai, H. Vindana,
K. Vencatasamy, K.V. Lawler, A. Salamat, R.P. Dias. Nature {\bf 586}, 373 (2020)
\bibitem{Scal}D.J. Scalapino. In ``Superconductivity'', p. 449, Ed. by R.D.
Parks, Marcel Dekker, NY, 1969
\bibitem{All}P.B. Allen, B. Mitrovi{\'c}. Solid State Physics, Vol. Vol. 37
(Eds. F. Seitz, D. Turnbull, H. Ehrenreich), Academic Press, NY, 1982, p. 1
\bibitem{AD}P.B. Allen, R.C. Dynes. Phys. Rev. {\bf 12}, 905 (1975)
\bibitem{Mig}A.B. Migdal. Zh. Eksp. Teor. Fiz. {\bf 34}, 1438 (1958) [Sov. Phys. JETP {\bf 7},
996 (1958)]
\bibitem{Sad_18}M.V. Sadovskii. Zh. Eksp. Teor. Fiz. {\bf 155}, 527 (2019) 
[JETP {\bf 128}, 455 (2019)]
\bibitem{Sad_19}M.V. Sadovskii. Pis'ma Zh. Eksp. Teor. Fiz. {\bf 109}, 165 (2019)
[JETP Letters {\bf 109}, 166 (2019)]
\bibitem{Sad_20}M.V. Sadovskii. Journal of Superconductivity and Novel Magnetism
{\bf 33}, 19 (2020)
\bibitem{Ikeda}M.A. Ikeda, A. Ogasawara, M. Sugihara. Phys. Lett. A {\bf 170}.
319 (1992)
\bibitem{UFN}M.V. Sadovskii. Usp. Fiz. Nauk {\bf 186}, 1035 (2016) [Physics Uspekhi
{\bf 59}, 947 (2016)]
\bibitem{Gork_1}L.P. Gor'kov. Phys. Rev. B{\bf 93}, 054517 (2016)
\bibitem{Gork_2}L.P. Gor'kov. Phys. Rev. B{\bf 93}, 060507 (2016)
\bibitem{Gork_3}L.P. Gor'kov. PNAS {\bf 113}, 4646 (2016)

\end{thebibliography}
\end{document}